\documentstyle[aps]{revtex}
\input epsf
\begin{document}
\tighten
\title{Taming Asymptotic Strength}

\author{Charanjit S. Aulakh$^{(0) (1), (2)}$}
\vspace{.3 cm}
\address{$^{(1)}$ {\it Dept. of Physics, Panjab University,
Chandigarh, India  }}
\address{$^{(2)}${\it International Center for Theoretical Physics,
Trieste, Italy }}

\maketitle
\begin{abstract}

Using a simple Asymptotically Strong  N=1 Susy SU(2) Gauge theory coupled to
a {\bf{5}}-plet  Chiral superfield
 we demonstrate the plausibility of the `` truly minimal ''  Asymptotically Strong  Grand
Unification  scenario proposed by us recently. Assuming a
dynamical superpotential consistent with the symmetries and
anomalies is actually induced non-perturbatively we show  the
gauge symmetry is  dynamically and spontaneously broken at a UV  scale
that is generically {\it{exponentially}} larger than the scales characteristic of  the
effective low energy theory.  The pattern of condensates in a semi-realistic $SU(5)$ ASGUT is
analyzed using the Konishi anomaly assuming condensation occurs as
shown by our toy model. The necessarily complementary relation of
ASGUTs to the Dual Unification program and the novel cosmogony
implied by their unusual thermodynamic properties are also
discussed.

\footnotetext{Email: aulakh@ictp.trieste.it (till 27 Oct.
 2002); aulakh@pu.ac.in (after 27 Oct. 2002).}

\end{abstract}

\vspace{0.3cm}

\section {  Introduction}

In a recent feuilleton \cite{trmin} we   agitated for  serious
 consideration of the possibility that the Grand Unification Gauge group
becomes strongly coupled  above the scale  of perturbative
unification ($M_U$). This behaviour is inevitable in any GUT
containing Higgs fields capable of giving realistic fermion mass
relations (FM Higgs) at tree level \cite{realmf1,realmf2}  since
the Casimir indices of such representations  are generically large
and hence drive the gauge coupling into the strongly coupled
regime  above their mass thresholds (the problem is especially
acute in SO(10) Susy GUTs that employ {\bf{126}}-plets but the
behaviour is generic ). We argued that the almost superstitious
avoidance of Asymptotically Strong (AS) models in the literature
was  unjustified. The  intuitive picture that emerges when the
basic reversal of roles between the  perturbative gauge
non-singlet and the confined gauge singlet  phases is accounted
for, is quite appealing  and does not violate any canon. Rather it
raises the possibility of quite novel solutions and insights into
the hoary puzzles of unification  physics. Even if unrealized in
nature it would constitute a remarkable logical completion that
needs no banishment to the realm of the unthinkable.

Reversing the usual logic that sees fermion masses  as a
peripheral or second order issue relative to the primary fact of
symmetry breaking we argued that their necessity in an important
component of the  inner rationale for symmetry breaking  in the
Standard  Model (SM) and GUTs. The SM ensures an observable
electromagnetic U(1)  gauge symmetry only by ensuring that all
charged particles  are in fact massive.  If this were not so then
the massless charged particles  would completely screen the
U(1) charge at low energies thus rendering it unobservable.
Moreover there are  other pathologies associated with massless
charged particles (of any spin)  in both Quantum and Classical
dynamics and this makes it all the more plausible that ensuring
their absence in the SM via the formation the Higgs Condensate is
a rationale behind the formation of the Higgs Condensate itself .
The  SM  fermion mass spectrum  is thus an important clue for the
choice of Higgs representations and so to the mode of symmetry
breaking of the GUT in which  the SM is embedded. In a truly
minimal (TM) model the low energy fermion mass and gauge coupling
data and the Fermion Mass (FM) Higgs representations needed to fit
it should determine the pattern and modality of unification
completely. In our proposal it is the low energy data itself that
forces us to take seriously the possibility of strong gauge
couplings above the GUT scale. Else we must  abandon painstakingly
developed and  otherwise sensible (Susy) GUTs e.g \cite{abmrs}
where the presence of $\bf{126 + {\overline{126}}}$ is used to
implement see-saw mechanisms for neutrino mass in an SO(10) Susy
GUT while preserving R-parity to low energies. The Casimir index
of these two Higgs multiplets alone is thrice that of the gauge
fields !

  Furthermore,  the presence of effectively
exact  Supersymmetry at the GUT scale ($\Lambda_U>> M_S$)  fairly
cries out for exploitation, given  all the attendant
simplifications. This is in  sharp contrast to the case of
``realistic'' SQCD where the supersymmetry breaking scale
completely dominates the condensation scale ($M_S >>
\Lambda_{QCD}$) and relegates the conclusions derived from Susy
QCD largely to the  realm of the qualitative \cite{rsqcd} and the
inapplicable. As we have shown elsewhere \cite{abs,ams,amrs,abmrs}
Susy  makes the analysis of symmetry breaking patterns and
phenomenology tractable even in very complex models. The enhanced
analyzability is conferred by the special holomorphicity and non
renormalization properties of  Supersymmetric theories
\cite{venyank,akmrv,ads,sei1}. Thus  N=1 strongly coupled Susy GUTs  are
prime candidates for application of the techniques developed for
the analysis of phase portraits of Susy Gauge theories
\cite{venyank,ads,sei1,reviews,pesk}. Unfortunately,
Asymptotically Strong (AS) models have received at most passing
mention in the enormous literature devoted to the dynamics of Susy
gauge theories. From time to time,  the possibility of AS Unified
theories (albeit with considerably larger gauge groups than
standard GUTs)  has been  resurrected \cite{strongun} but no
particular merit of uniqueness or  genericity  has been available
to justify the extensions proposed .  Moreover it was 
unclear how any significant step beyond the bald statement of UV
strength could be made about the UV dynamics of such theories and
its relevance to the structure of the low energy theory.

In \cite{trmin} we proposed that  the (scale inverted) analogy of
AS theories with familiar strong coupling QCD dynamics suggests  a
very appealing intuitive picture of the effect of AS.
  Since the low energy and long distance
theory is weakly coupled the principal effects of the strong coupling in the UV are likely to
be in the formation of condensates that break gauge symmetry and thus define
the vacuum of the  perturbative effective gauge theory at larger length scales
$>>\Lambda_U^{-1}$. Such condensates would also directly modify
our picture of the nature of elementary particles at the smallest scales where the gauge flux
would be strongly self attracting. This picture also suggested a close
connection with Dual unification \cite{vachas} in
which matter elementary particles are
seen as monopole solitons of a dual gauge theory. Indeed the dual unification picture becomes
much more coherent, consistent and plausible
when seen as the dual of an AS Susy GUT; as
indeed is {\it{required}} by its very assumptions.

For these reasons we think it necessary and possible to investigate the  scenario in which FM
Higgs residuals with mass thresholds just above the scale of perturbative unification
($M_U$)  drive the gauge coupling into the strong coupling regime at a scale
$ \Lambda_{U}$
which is quite close to $M_U$ in realistic models. The energy ranges that need to be
characterized divide naturally into  three : {\bf{I)\ }} the UV strong coupled regime
($ E > \Lambda_U $) {\bf{II)}} the perturbatively coupled supersymmetric intermediate regime $
\Lambda_U >  E > M_S$ and {\bf{III)}} the  standard low energy  MSSM/SM regime
 below the supersymmetry breaking scale : $M_S>E $.
We shall take the pragmatic epistemic position that the
theoretical task is no more than to construct economical and
minimal models  within which each of these qualitatively distinct
regimes can be quantitatively characterized and their dynamics
{\it{understood}} via simple ``pictures'' of the processes
involved. The relation and translation between the effective
theories that characterize each regime, and their derivation from the underlying
fundamental gauge theory,  is $-$ as
familiar from QCD and Chiral Perturbation theory $-$ expectedly a
difficult problem that we may expect to resolve only
semi-quantitatively in an analytic context.

We argue that the novel interpretative context of  UV condensation
(regime {\bf{I}}) and GUT scale to QCD scale  liberation of gauge
variant degrees of freedom(regime {\bf{II}}) makes available new -
and pleasing - possibilities that can further our  insight into
the nature of elementarity and unification. For clarity and
definiteness we choose a  simple renormalizable AS model
consisting of N=1 Susy SU(2) Gauge theory coupled to a Chiral
multiplet transforming as  a {\bf{5}} dimensional SU(2) irrep and
ask what are the effective theories that characterize its
behaviour in the  regimes described above. We make the plausible
assumption that in the crucial intermediate energy perturbative
regime {\bf{II}} the relevant degrees of freedom are the usual
gauge non-singlet ones described by the N=1 Susy YM lagrangian
with a  renormalizable tree level superpotential $W_t$ {\it{and a
dynamically induced superpotential arising from non-perturbative
dynamics in a manner consistent with holomorphicity, anomalies and
 global (super)symmetries in complete analogy with the well established dynamical
superpotentials that arise in the AF case}} \cite{venyank,ads,sei1,reviews,pesk}. In
region {\bf{I}} we take the relevant degrees of freedom
to be  (`t Hooft anomaly matched )  gauge singlet chiral moduli interacting via the same
dynamical superpotential as in the perturbative regime but differing in (at least) an
independent and unknown Kahler potential.
We assume that the deviations from the canonical form of
 the Kahler potential (and the gauge kinetic chiral functions) in the perturbative region
{\bf{II}} are negligible (see Section II). This ansatz is in accord with the standard picture
of confinement enforced on non-gauge particles by singular wave function renormalization, and,
correspondingly, the divergence of their renormalized effective masses.
In the perturbative region {\bf{II}} we can extract
significant information using the  definition of the moduli as polynomial composites
in the chiral multiplets. Region {\bf{III}} or the Susy EW breaking/QCD confining regime
is governed by the standard MSSM/SM dynamics.

The plan of the paper is as follows. In Section II we set out our
scenario and intuitive picture of how an AS GUT  works. In Section
III we define and discuss the strongly coupled SU(2) Gauge plus
quintet Higgs model referred to above and obtain the possible vevs
of the Higgs fields . In Section IV we discuss  the pattern of
dynamical symmetry breaking (DSB) and gaugino condensation implied
by the vevs of Section III and the constraints on the dynamical
superpotential that may be inferred using the Konishi
Anomaly\cite{kona} and decoupling arguments. In Section V we
discuss  the lessons and implications of our simple model for the
project of constructing viable realistic  AS GUTs.
 We conclude in Section VI  with a discussion of the  appealing
relation ASGUTs necessarily and naturally enjoy with  the
Dual unification picture \cite{vachas}
in which the elementary  particles of the SM appear as monopoles of a
dual gauge group and how it resolves otherwise problematic features of those models. We
thus motivate a  new, supersymmetric, version of the dual unification scenario\cite{wip}.
Finally we discuss  the ASGUT novel cosmogony  implied by  {\it{liberation}} of degrees of
freedom by {\it{cooling}} in ASGUT thermodynamics.

\section{ AS Dynamical Scenario}

A plausible picture of the dynamics of AS theories may be drawn using the
analogy with the known features of IR strong gauge theories.
The reversal of the roles of small and large length scales in the case
of AS theories  turns the usual picture of confinement ``inside out''.
It is this conceptual leap ,  diametrically opposed yet closely analogous to the inbred
picture of QCD and its effective low energy theory of colour singlet hadrons that
jumps the barrier to realizing that AS theories can be an
equally consistent ``other half'' of the space of Yang Mills theories.

When the gauge coupling becomes strong one  expects  gauge singlet
condensates (in particular the gaugino condensate  familiar from Susy QCD)
to form and to be characterized by a scale $\Lambda_{U}$ ($\equiv \Lambda $)
at which the (one-loop) RG equations indicate a divergence in the gauge coupling .
Above $\Lambda $ the  gauge non-singlet degrees of freedom become confined due to
singular wavefunction renormalizations which give them infinite effective masses. Thus
the residual dynamics must involve suitable gauge singlet degrees of freedom.
 It is natural to choose the D-flat moduli which are so central in Asymptotically Free (AF)
Susy theories for the leading  role in the AS UV dynamics as well since their
masses are less than the typical energies  in region {\bf{I}}.
As in the AF case the 't Hooft anomaly matching conditions\cite{hoof}  serve
as a valuable cross check of the completeness and consistency of the set of modes retained.

The electric permeability of the UV  condensates must exhibit the
peculiar inversion referred to above :  self attraction of  a  point particle's
electric flux (equivalently strong screening by the condensate) implies that
it should  form a gauge singlet tangle or ball of size
$\sim \Lambda^{-1}$ surrounding the putative
point charge. This tangle will not be resolvable by any gauge non singlet probe: {\it{thus
conferring an  elementary size on fundamental gauge non-singlet particles}}. On larger
scales however the electric flux is weakly coupled and hence the
 gauge charge  of  an elementary particle will  appear as the source  of flux
streaming out freely (on scales $>>l=\Lambda^{-1}$)  from a  gauge
singlet core (see Fig.1) characterized by the gauge charge as a
``surface '' parameter (somewhat like the quantum numbers of a
black hole in the horizon-membrane picture \cite{bhmem}).  Such a
structure could  be described as a soliton or bag of the  Chiral
gauge singlet dynamics used to characterize the  high energy
phase of the theory married to the perturbative  gauge theory at
the core boundary .

\begin{figure}
\centerline{\epsffile{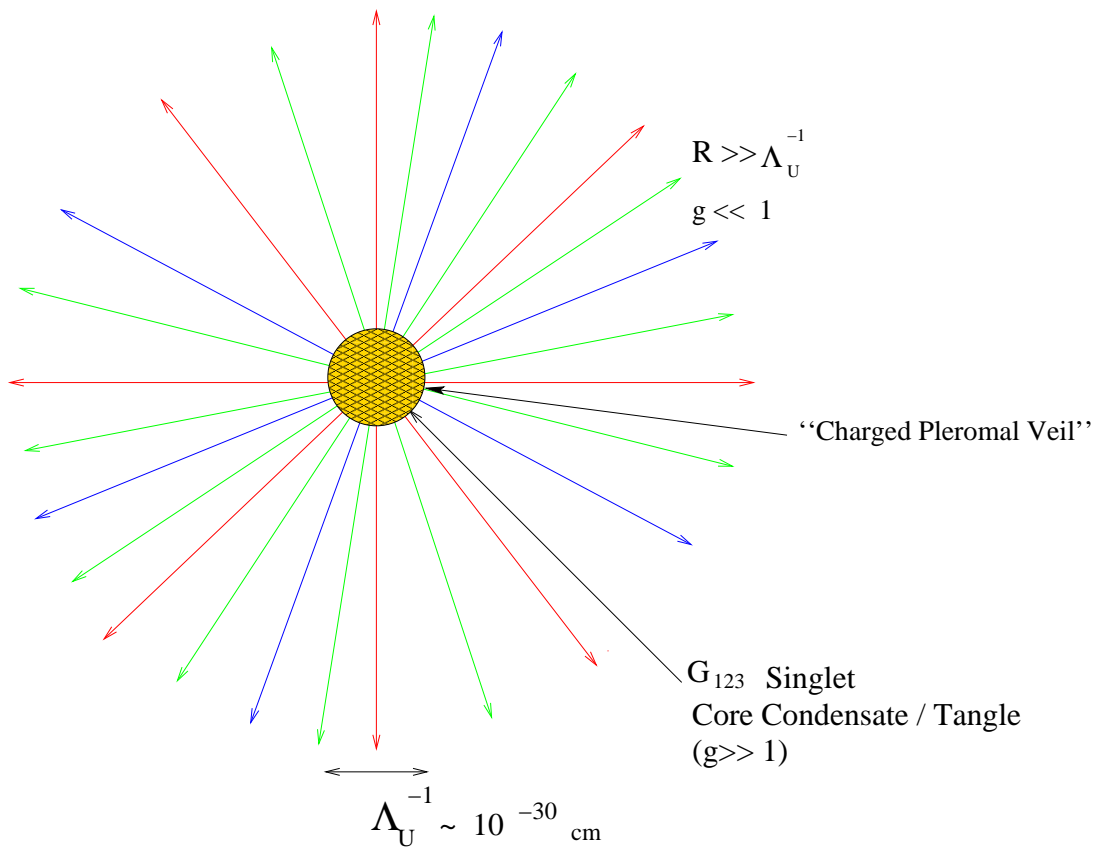}}
\caption{The Pleromal Heart of a Quark : The interior of the core condensate cannot be probed 
by charged probes and the quark charge is a surface parameter of the core tangle.}
\label{phqk}
\end{figure}

Since the gauge coupling is small below $M_U$
the low energy dynamics is  a perturbative theory of gauge non-singlet particles
interacting with the massless gauge bosons of the gauge symmetry left unbroken by the
condensation at the UV scales.  This is in sharp contrast to QCD where the perturbative
dynamics is indifferent to the confining condensates precisely because the energy scales of
 the perturbative theory are much {\it{larger}} than those of the QCD condensates
($\Lambda_{QCD}$  functions as an IR cutoff).  Our prescription
for unravelling the very novel features presented by AS UV
condensation is thus to replace the fundamental Gauge theory by
effective field theories  in each of  the  dynamical regions. For
region {\bf{I}} ( energy scales $E>\Lambda_U $) we propose  a
theory of the  anomaly matched gauge singlet Chiral moduli ($X_{(n)}$) of the
fundamental N=1 Susy Gauge model interacting according to a
dynamical superpotential $W_d(X_{(n)},\Lambda,m,\lambda)$  (in
addition there is  the  tree superpotential of the fundamental
theory which depends on the masses and couplings
$m_i,\lambda_i^{(n)}$ of the Chiral multiplets $\Phi_{(i)}$)
which represents the strong coupling dynamics. The Kahler potential in region {\bf{I}} 
$K_I(X,X^*)$ is a highly non perturbative object representing the condensation 
of the moduli theory and the confinement of the gauge and gauge variant chiral modes. However 
one could still apply  supersymmetric chiral perturbation theory techniques to it
in direct analogy with the AF case \cite{rsqcd}. Many of the
features of $W_d$ can be deduced or constrained on the basis of
(super)symmetry, holomorphicity and decoupling requirements
\cite{reviews}. In the perturbative regime {\bf{II}}, $\Lambda_U > E > M_S $, we have  the
{\it{effective  spontaneously broken}} gauge theory  of modes left light by
the UV condensation of the gauge singlet moduli. Such a theory will be described by three 
functions : the Kahler potential $K_{II}(\Phi,\Phi^*)$, the superpotential $W(\Phi) = W_d +
W_t$ and  the chiral gauge kinetic coupling function $f^{ab}_{II}(\Phi ) $.
In this paper we shall focus on Region {\bf{II}} and the 
effects coded in the superpotential $W_d$ assuming the Kahler potential and gauge kinetic
function are effectively canonical. Note however that while a canonical Kahler potential 
for the perturbative modes  is a reasonable assumption , the effects 
of a non-canonical gauge kinetic 
function can be critical for issues of supersymmetry breaking in the presence of 
gravity \cite{nilles}.
Indeed it is very interesting to study\cite{wip} the generation of non-canonical 
structure in this function in $N=2$ AS theories 
(or perturbations thereof) where one can seek to apply the 
techniques developed \cite{seiwit} for N=2 AF theories.

  An important consistency check on the set of gauge singlet moduli retained
in the UV effective theory of region {\bf{I}} is that they should satisfy the 't Hooft anomaly
matching conditions \cite{hoof,reviews,pesk} for the  unbroken global symmetries of the
model with
respect to the anomalies calculated using the gauge non singlet or fundamental spectrum.
In IR strong models this set is found to exclude the glueball chiral
superfield $ S=(1/32\pi^2) W^a_{\alpha} W^{a\alpha} $ (where $W^a_{\alpha}$ is
the (bare)  chiral gauge supermultiplet in `holomorphic'
normalization\cite{shifh,arkmur,kon2}. We also find that the anomaly matching conditions
exclude S from the   singlet spectrum for region I. As in the IR  strong case, S plays a
vital role in  understanding the strong coupling condensates since the Konishi anomaly
\cite{kona}  furnishes the value of  the gluino condensate (i.e the scalar component of $S$)
in terms of the expectation value of the chiral condensates .
In the initial studies \cite{venyank}  low energy
effective actions were built including this glueball field. Then  one can
code  in the anomalies of the theory into the effective Lagrangian since the Glueball
superfield $S$  contains $F^a_{\mu\nu}{\tilde F}_a^{\mu\nu}$
in its auxiliary component.  However it was argued \cite{sei1}  that the
effective low energy action can lead to misleading and ambiguous answers if
only a partial set of the ``heavy'' degrees of freedom is included in it.
 This caveat is supported by the fact that inclusion of the glueball field S along
with the singlet moduli in the effective theory typically violates the 't Hooft
anomaly matching conditions for the unbroken global symmetries.  On the other hand
 it is not clear that inclusion of  a finite set of additional composite fields could
not cancel the contribution of the glueball field. Then this additional anomaly matched set,
being qualitatively lighter than the coloured fields which have 
receded to infinite running mass, could play a useful
role in describing the  dynamics of modes with masses
less than or near to the confinement scale or even of massless modes that could appear
as bound states . In any case the action with $S$ can serve the useful functions described
above even if it is not trustworthy for quantitative dynamics.

The  Konishi Anomaly relations\cite{kona} are superfield $U(1)_c$ Ward-Takahashi identities
that arise when   the non-invariance of the path-integral measure under an arbitrary chiral
superfield rephasing of {\it{individual}} chiral superfields $\Phi_{(i)}$  is taken into
account. In the absence of sources it reads ($V_i$ is the gauge vector superfield)

\begin{equation}
0= \Big\langle -{{\bar D}^2\over 4}{\bar \Phi^{(i)}} exp[V_i]\Phi_{(i)} +
\Phi_{(i)}{{\partial W_{tree}}\over {\partial\Phi_{(i)}}}  + {{2T_{(i)}}\over {32\pi^2}}
W^{a\alpha} W^a_{\alpha} \Big\rangle
\label{kar0}
\end{equation}

The lowest component of the Konishi anomaly yields a very useful set of relations involving
the  gaugino condensate  $\langle S|_{\theta =0}\rangle =\langle(-
\lambda^a\lambda^a/32\pi^2)\rangle $
and vevs of the  different invariants formed from chiral superfield scalar components
corresponding to the chiral  invariants occurring in the tree superpotential $W_{tree}$:

\begin{equation} \langle S \rangle = - {1\over{2T_{(i)}}} \Big\langle \phi_{(i)}{{\partial
W_{tree}}\over {\partial\phi_{(i)}}} + {1\over{2\sqrt 2}}\{{\bar Q}_{\dot\alpha},{\bar
\psi}^{\dot\alpha i}\phi_i\}\Big\rangle
\label{kar}
\end{equation}
There is no sum over the index $ (i)$, $T_{(i)}$ is the index of
$(i)$th representation in the normalization where the fundamental is $1/2$,
and $W_{tree}$ is the tree level superpotential
of the underlying gauge theory. The vev of the anticommutator with the
supercharge ${\bar Q}_{\dot\alpha}$ above
vanishes to the accuracy that supersymmetry is preserved. On dimensional
grounds its value can be tentatively estimated as $M_S^3$ or perhaps as
$\sim M_S^2 \langle \phi_i\rangle $. A clearer understanding of its role can come
about only once the  source of supersymmetry breaking in the theory is specified
(see Section 6) and the effect of  soft supersymmetry breaking terms on the derivation of
the Konishi Identities eqn.(\ref{kar0}) determined.
In eqn.(\ref{kar}), in the limit of exact supersymmetry,  the glueball field S is not
renormalized perturbatively and nor is $W_{tree} $ or the operators
$\Phi_{(i)} \partial/\partial\Phi_{(i)} $.
Thus we may expect that this relation is RG invariant and should be respected at
all scales to the accuracy to which supersymmetry is preserved. On the other hand, since a
realistic model will have soft Susy breaking masses that violate Susy explicitly in the
Lagrangian , the assumptions of the derivation of eqns(\ref{kar0},\ref{kar}) will not hold
and thus, at least till we have better control on the mechanisms of Susy breaking in ASGUTs,
we should not trust these relations to an accuracy better than $M_S^3$ or perhaps
$M_S^2\langle\phi\rangle$ .

These equations thus impose strong
restrictions and requirements  on the pattern of condensates in the theory.
  In the presence of a tree superpotential we expect that gaugino condensation at a
scale $\Lambda$ implies UV condensation of moduli (and relations
among different moduli vevs) and vice versa at least as long as
supersymmetry  is unbroken. In the perturbative theory the moduli are defined as
polynomials in the gauge variant chiral fields and their (large)
vevs determine the broken symmetry  of the low energy gauge theory
while characterizing it in a gauge invariant way. A subtle but
inescapable possibility  in this regard is that of {\it{quantum}}
phase transitions in which e.g the product of  two  Higgs fields
obtains an expectation value but {\it{not}} the fields themselves.
This is a real possibility since the scale of perturbative
unification $M_U$ is sufficiently close to the UV strong coupling
scale $\Lambda_U$ to make quantum condensates  important. However
we shall resist the temptation of invoking such condensates, until
and unless it is unavoidable, so as to retain maximal
calculability in the effective gauge theory below the GUT scale.
The effective low energy field theory must thus be shown to be
compatible with the these very novel quantum vevs of gauge singlet
composite operators which preserve all the symmetries of the
theory, by definition.

\section{SU(2) AS model}

Our model consists of a N=1 Susy  SU(2) Gauge theory coupled to a SU(2) {\bf{5}}-plet
($SU(2)$ spin $j=2$) Chiral superfield  presented
as a symmetric traceless $3\times 3 $ matrix: $\Phi_{ij}
=\Phi_{ji}; \Phi_{ii}=0 ; i,j=1,2,3 $ .
The additional feature relative to the
familiar adjoint representation is that such a complex symmetric matrix cannot,
in general, be diagonalized by a complex orthogonal transformation.
However one observes that firstly the (upper triangular) Jordan canonical form of $\Phi$ is

\begin{equation}
\Phi = S \left(\begin{array}{ccc}
a & c &d \\
0&b&e\\
0&0&-(a+b)
\end{array}\right) S^{-1}
\end{equation}

\noindent so that $X_{(n)} = Tr \Phi^n = a^n + b^n +(-(a+b))^n $
are  dependent on only two independent complex numbers $a,b$ and as such
only two of them, say the quadratic and
cubic invariants $X_{(2)}=X, X_{(3)}=Y$, need be considered independent.

This can also be seen from the less familiar result
that complex orthogonal transformation by $O_c$ can
be used to put \cite{craven} a complex
symmetric matrix in a block diagonal canonical form :

\begin{equation}
\Phi= O_c Diag(B_1,B_2,.....) O_c^T
\label{canon}
\end{equation}

\noindent where a diagonal block $B_i$ is either a $1\times 1$
submatrix consisting of an eigenvalue of $\Phi $ or a $q\times q$
symmetric submatrix of form $\lambda_i I +E_i$ where $\lambda_i$
is an eigenvalue and $E_i$ has all its eigenvalues zero, all its
eigenvectors quasi-null and an eigenspace of dimension $d\leq q$.
The matrices $E_i$ arise because in general a complex symmetric
matrix can have invariant eigen-subspaces of zero Euclidean
length. However in the present instance  the additional freedom
associated with these subspaces is irrelevant to definition of the
gauge singlet moduli and also to the extremization of the
superpotential w.r.t the field $\Phi$.  The D-terms can be written
as a commutator $[\Phi,\Phi^\dagger]$ . After fixing the $SU(2)_c$
gauge freedom by putting $\Phi$ in the canonical form
(\ref{canon}) one sees that the pseudo-moduli contained in $E_i$
must vanish so that one can work with the diagonal form .

The index of the {\bf{5}}-plet (in the normalization where the fundamental has index
$1/2$) is $10$ . Thus this theory has a one loop gauge beta function
$\beta(g) =- b_0 g^3/16\pi^2 ; b_0=3\times 2- 10=-4$ and
 hence its coupling grows in the ultraviolet.
In the absence of a superpotential the underlying theory has $U(1)_{\Phi}$  symmetry
under which the chiral superfield $\Phi$ has charge 1 and in addition a
 $U(1)_R$ R-symmetry
under which the Grassman coordinate $\theta_{\alpha} $  has charge
1 , the gaugino  -1 and the components $(A,\psi,F)$  of $\Phi$
have  charges (0,1,2) respectively. Both these symmetries have
mixed $ U(1) SU(2)^2 $ anomalies but the linear combination

\begin{equation}
{\widehat Q} = {1\over 2} (\sum_i T(i) (q^0_i Q_{\Phi} - q_i^{\Phi} Q_0) )
 = 4 Q_{\Phi} - 5 Q_0
\end{equation}

\noindent generates a  non anomalous U(1) R symmetry. The charges of the various Chiral
superfields
and components (and of the holomorphic parameters introduced below)
are given in Table I.


\begin{center}
\begin{tabular}{l|rrr}
Field\hspace{.3 cm} &\hspace{.3 cm}$Q_{\Phi}$ &\hspace{.3 cm}$Q_{0}$& \hspace{.3 cm}
${\widehat Q} $ \\
\hline
$\theta_{\alpha}$ & 0 & 1 & -5\\
$\Phi$ &1 & 0 &4 \\
$\psi_{\Phi}$ &1 &1 & -1\\
$\lambda_{\alpha}$ &0&-1&5\\
$S$ &0 &-2 & 10\\
$m$ &-2 &-2 & 2\\
$\lambda$ &-3 &-2 &-2\\
$\Lambda^{-4}$ &20 &16 & 0\\
$\Lambda_d^{3}$ &0 &-2 & 10\\
\end{tabular}
\end{center}
{\bf Table 1 :}$\quad U(1)$  charges of fields and parameters.

The contributions of the elementary fermions to the ${\widehat{U(1)}}^3$ and
${\widehat{U(1)}}$ (gravitational) anomalies are $370 = 3\cdot 5^3 + 5\cdot (-1)^3 $ and
$3\cdot 5 + 5\cdot (-1) =10$ respectively . The fermionic components of the moduli fields
i.e $\psi_{X,Y}$  have ${\widehat{U(1)}}$ charges 3,7 and hence precisely match these
anomalies . The glueball field
$S = (32\pi^2)^{-1} W^a_{\alpha}  W^a_{\alpha} $    cannot
therefore be consistently included alone in the relevant degrees of freedom.

The most general  renormalizable tree superpotential is

\begin{equation}
W_{t} = - {m\over 2} X  +{ \lambda\over 3} Y
\end{equation}

Prima facie this potential violates all the $U(1)$ symmetries .
However by considering\cite{sei1} the parameters $m,\lambda$ as
expectation values of chiral fields one can exploit the
holomorphicity of the Susy theories w.r.t chiral fields and hence
we assign the charges shown in Table I to $m$ and $\lambda$ so as
to formally preserve the $U(1)$ symmetries.

The one loop gauge coupling evolution in the range  $\Lambda_U > \mu > m $ is

\begin{equation}
{1\over {g^2(\mu)}} = {1\over {g^2(m)}} - {4\over {8\pi^2}}ln {\mu\over m}
\end{equation}

If the Higgs field decouples completely and leaves an unbroken  pure $SU(2)$ 
gauge theory below the
scale $m$ the in the range  $m > \mu > \Lambda_d $ the gauge coupling evolves as
\begin{equation}
{1\over {g^2(\mu)}} = {1\over {g^2(m)}} + {6\over {8\pi^2}}ln {\mu\over m}
\end{equation}

Here  $\Lambda_U$ is the scale where the (one-loop)
running coupling of the AS model diverges (henceforth abbreviated to $\Lambda$) and
$\Lambda_d$ the corresponding scale for the
effective theory in which the Higgs multiplet has decoupled completely without breaking the
symmetry. It can also happen that the Higgs multiplet breaks the symmetry partially : i.e in
this case down to U(1). In that  case the low energy theory will be a pure $U(1)$
gauge theory coupled to charged Higgs remnants (if any)
and the $U(1) $ gauge coupling evolves accordingly
and must be matched at the scale $m$ to the coupling in the high energy theory.
However, in more complicated models with larger gauge groups,  the low energy symmetry
can be non-abelian. Let $m_t$ be a notional low energy
scale  where the gauge coupling of the low energy effective theory is specified , then
by matching gauge couplings of the UV and IR theories at the threshold $m$ (in the DR
scheme)\cite{venyank,ads,reviews,finpou} one gets

\begin{eqnarray}
\Lambda_d &=& m_t e^{({-{4\pi^2}\over {3 g^2(m_t)} })} =
m e^{({-{4\pi^2}\over {3g^2(m)}})}\nonumber \\
\Lambda &=& m e^{({2\pi^2\over { g^2(m)} })} =({{m^5}\over \Lambda_d^3})^{1\over 2}  \\
\epsilon &=& {m\over \Lambda} = ({{\Lambda_d}\over m})^{3\over
2}=({{\Lambda_d}\over {\Lambda}})^{3\over 5}\nonumber
\end{eqnarray}

It is important to note that the decoupling parameter $\epsilon $ must be taken
to zero as $m,\Lambda\rightarrow \infty $ while keeping $\Lambda_d$ constant for consistency.

  When building effective superpotentials invariant under
all symmetries (anomalous or not) it is also useful
to assign $\Lambda^{b_0} $ the charges 20,16,0  and use $\Lambda$ to adjust dimensions in
dynamical expressions\cite{reviews,pesk}.
Proceeding in the familiar way \cite{reviews} one finds that the symmetries of the theory
restrict the dynamical superpotential to the form

\begin{eqnarray}
W_d &=& X^{5\over 4}\Lambda^{1\over 2} {\widehat W (v,w,z)} +  \Lambda_d^3
f({{m\lambda}\over \Lambda})\\
v &=& X^3/6Y^2 \qquad w = m/ X^{1\over 4}\Lambda^{1\over 2} \qquad z=\lambda  X^{1\over4}
/\Lambda^{1\over 2} \nonumber
\end{eqnarray}
 Note that the exponent of $ \Lambda $ ($T=10$ is the index of the
Higgs, N=2 the gauge Casimir) is  $(3N-T)/(N-T)$ and thus has the form familiar from QCD but
is positive due to a
double negative (numerator and denominator). As is well known, these exponents
are determined by symmetry and dimensional analysis coupled with the 
renormalization group \cite{venyank,akmrv,reviews}.
The unknown function $\widehat W$ is a power series

\begin{equation}
{\widehat W} = \sum_{\alpha,\beta} {\widehat{W}}_{\alpha,\beta} (v) w^{\alpha} z^{\beta}
=\sum {\widehat{W}}_{\alpha,\beta}(v) m^{\alpha}\lambda^{\beta}
\Lambda^{-{{(\alpha+\beta)}\over 2}}
X^{{\beta -\alpha}\over 4}
\label{what}\end{equation}

As in the case of QCD such a function can, presumably, arise via (fractional) instanton
or other non-perturbative effects. Our basic assumption and approximation is that $W_d$
is non zero and admits an expansion in the perturbing parameters $m,\lambda $.
As we shall see, the presence of such a function, which respects all symmetries and anomalies,
allows one to implement the demands of decoupling in a straightforward manner and presents no
difficulties of the sort one might expect to encounter if  it was forbidden
for deep  reasons. On the contrary, we show that the consistency of the physical
picture requires a dynamical superpotential, or at least a constraint
 implementing condensation.  It is perhaps worth noting that even in the
well studied case of Susy QCD the amplitude of the
 induced superpotential is only reliably
calculable\cite{finpou,hol}, so far,  in
the case $N_f=N_c -1$ while the amplitudes in all cases where fractional powers of
$\Lambda^{b_0}$ enter are deduced only on the basis of decoupling.
Note that if the function $\widehat W$ respects the form of the (fractional) instanton
expansion then (schematically)

\begin{equation}
W_d \sim (\sum_{n=1} \Lambda^{n b_0} )^{{1\over {(N-T)}}}
\sim \Lambda^{b_0\over{N-T}} (1 + {1\over {N-T}} \Lambda^{b_0} + ....)\sim\Lambda^{1\over
2}(1 +\Lambda^{-4} + ..)
\end{equation}
Where we have written the instanton suppression factor $e^{ -8\pi^2/g^2} $ as
$ \Lambda^{b_0}$ using the one loop definition of $\Lambda$. Thus in eqn.(\ref{what})
 $\alpha +\beta = 8n , n= 0,1,2 ... $. Since we regard the superpotential
parameters as small perturbations to the main dynamics we see that to leading order in
$m, \lambda $ we can write

\begin{equation}
W ={\Lambda^3} [x^{5\over 4} G(v) -{\epsilon \over 2} x +
{\lambda\over 3} y]
\label{W}\end{equation}

\noindent where $x=X/\Lambda^2 , y =Y/\Lambda^3 , \epsilon= m/\Lambda =(\Lambda_d/m)^{3/2}$
are dimensionless and we have dropped the constant term as irrelevant. Our
assumption is thus that, at least in the perturbative regime, the
superpotential $W_d$  represents the main non-perturbative effects. Its  implications
for the vacuum structure can  be evaluated by extremizing it w.r.t the perturbative
field $\Phi$. The same result may also be obtained by demanding \cite{venyank} that a
superpotential dependent on $X,Y$ and the glueball field $S$  reproduce the ABJ
anomalies and be invariant under the $\widehat{U(1)}$ symmetry. In that case one finds that

\begin{equation}
W_d(X,Y,S) =  S [ln {{X^{10} \Lambda^4 H(v, {{mX}\over S} , {{\lambda X^{3\over 2}}\over
S}) }}] -8 S ln S
\end{equation}

\noindent satisfies all the conditions imposed even if $H$  is taken to be an arbitrary power
series in its arguments. From such a superpotential one could, given H, eliminate $X,Y$ using
the vacuum equations of motion ,  to obtain 
a superpotential $W(S,m,\lambda) $  which can be used to characterize the gaugino condensate
in the  vacuum of the low energy pure gauge theory.
In the process one finds, to leading order in $m,\lambda $ the Konishi anomaly relation
eqn(\ref{kar}).  Conversely, in the same approximation, if one
eliminates the field $S$ in favour of $X,Y$ one finds
$W_d=8\langle S\rangle (X,Y)$.   The Konishi  relation
eqn.(\ref{kar}) between gaugino ($S$) and chiral multiplet ($X,Y$)
condensates is then readily seen to follow from from  the
extremization of the total superpotential. The effects of the
$m,\lambda$ dependence of $H$ appear as corrections.  Recently
remarkable claims have appeared \cite{vafdca} regarding the
calculability of the full ($m,\lambda$) corrections to $W(S)$ and
$\langle W\rangle (m,\lambda) $  via large $N_c$  matrix models
(at least in theories where the chiral multiplets can be written
as matrix representations of the gauge group). Such results would
permit evaluation of the chiral condensates and (via a Legendre
transform) construction of $W_d$ . Moreover the Konishi relation
eqn(\ref{kar}) is a rigorous relation that should be obeyed to all
orders in the parameters $m,\lambda$ and this will also constrain
$H$. Since we are here working only to leading order in
$m,\lambda$ the form of our superpotential is already consistent
with the Konishi Anomaly relation as explained. One gets

\begin{equation}
\langle S| \rangle  =- {1\over{20}} \sum_{A = 1}^{5}
\phi_A{\partial\over{\partial\phi_A}}W_t
={1\over{20}}(m X| -\lambda Y|)
\label{kar1}
\end{equation}
The $|$ refers to taking the scalar component and will be implicit hence forth.
Since the limit $\Phi\rightarrow 0$ should be nonsingular for AS theories, the behaviour of
the unknown
dynamical function $G(v)$ as $v\rightarrow 0,\infty $ is constrained. If
$\gamma_0,\gamma_{\infty} $ are the exponents of G in these limits then one obviously has
$\gamma_0 \geq -{5\over {12}}, \gamma_{\infty} \leq 0 $ since these represent the
conditions required to approach $X=0$ at constant $Y$, and $Y=0$ at constant $X$, respectively,
without $W_d$ becoming singular.
The vanishing of the $ F_{\Phi}$ terms  gives

\begin{equation}
2\Lambda {W}_x \Phi  + 3 {W}_y (\Phi^2 - {1\over 3} tr\Phi^2) = 0
\label{emphi}
\end{equation}

where
\begin{eqnarray}
{W}_x = x^{1\over 4}({5\over 4} G(v) + 3 v G'(v)) - {\epsilon \over 2}\nonumber \\
{W}_y =  x^{-{1\over 4}}(- 2{\sqrt{6}} v^{3\over 2} G'(v)) + {\lambda \over
3} \label{wxwy}
\end{eqnarray}

The solutions to these equations can be divided into

{\bf{A)}} Singlet mode solutions which arise when $W_x = W_y = 0$ .

{\bf{B)}} Perturbative mode solutions which arise when $W_x, W_y$ are both non zero.

\vspace{.3 cm}
{\bf{A) Singlet mode solutions}}
\vspace{.3 cm}

Multiplying the two equations in $W_x=W_y=0$ using eqns.(\ref{wxwy}) we get

\begin{equation}
({5\over 2} G(v) + 6 v G'(v)) (6v)^{3\over 2} G'(v)) = {{\epsilon  \lambda}}
=\eta
\label{singmod}\end{equation}
thus given $G(v)$ one can determine $\langle v\rangle$,  while the second of
eqns(\ref{wxwy}) fixes $X$ once the value of $v$ is found :

\begin{equation}
\langle x\rangle^{1\over 4}  =\langle {{(6v)^{3\over 2} G'(v)} \over \lambda } \rangle
\label{x1/4}\end{equation}

We now perturb in the small parameter $\eta$ . Three types of solutions can exist
depending on the solution $v_0$ of the equation

\begin{equation}
(5 G(v_0) + 12 v_0 G'(v_0)) (6v_0)^{3\over 2} G'(v_0)) = 0
\label{v0eq}\end{equation}

These are
\begin{itemize}
\item[{\bf{i)}}] $v_0$ non-zero and finite and determined by
\label{v0}
\end{itemize}
\begin{equation}
5 G(v_0) + 12 v_0 G'(v_0)\equiv 5 G_0 + 12 v_0 G_0'=0
\end{equation}
\begin{itemize}
\item[{\bf{ii)}}] $v_0 =0 \qquad\qquad $ {\bf{iii)}} $v_0 =\infty $.
\end{itemize}
\vspace{.3 cm}

The exponents and amplitudes of  perturbation in the small
parameter $\eta $ are defined as :

\begin{eqnarray}
\langle v \rangle &= v_0 + v_1 \eta^{\delta} + .....\nonumber \\
\langle x \rangle &= x_0 + x_1 \eta^{\theta}  + .......\\
\langle y \rangle &= y_0 + x_1 \eta^{\omega}  + .......\nonumber
\end{eqnarray}

\vspace{.3 cm}
{\bf{i)  $v_0\neq 0,\infty$  } }
\vspace{.3 cm}

When $v_0,G_0'$ are non zero one finds
\begin{eqnarray}
x_0 &=& (5{\sqrt{3v_0\over 2}} {G_0 \over\lambda})^4\nonumber \\
y_0 &=& {{x_0^{3\over 2}}\over {\sqrt{6v_0}}}
\end{eqnarray}

\noindent Continuing one can find the higher order corrections . Thus $\delta=\theta=\eta=1 $
and $v_1,x_1,y_1$ are determined in terms of $v_0$ and the derivatives of $G(v)$ at
$v_0$.  This type of solution generically gives

\begin{equation}  X \sim {\Lambda ^2 \over \lambda ^4}, \qquad Y\sim {\Lambda^3 \over
\lambda^6} \qquad i.e \qquad
\Phi\sim {\Lambda\over {\lambda^2}}
\end{equation}

\noindent and thus represents  dynamical breaking of the gauge
symmetry at the strong coupling scale $\Lambda$. The gauginos also
condense strongly (see the next section for a discussion). Clearly
in the absence of further information about $G(v)$ there is
nothing useful further to say.  In the other two cases the
solutions are governed by the exponents $\gamma_{0,\infty}$ of the
function $G(v)$ as $v\rightarrow 0,\infty$

\vspace{.3 cm}
{\bf{b) $\quad v_0=0$ }}
\vspace{.3 cm}

In this case the exponent $\delta$ must be positive for consistency .
Provided $\gamma_0\neq 0, -5/12 $ one finds

\begin{eqnarray}
\delta &=& {2\over {4\gamma_0 +1}} \Rightarrow \gamma_0 > -{1\over 4}\nonumber \\
\theta &=& {{4 + 8 \gamma_0}\over {1 + 4 \gamma_0}} \\
\omega &=& {{5 + 12 \gamma_0}\over {1 + 4 \gamma_0}} \nonumber
\end{eqnarray}

and

\begin{eqnarray}
v_1 = ({2 \over { 6^{3\over 2} \gamma_0 h_0^2 (5+ 12 \gamma_0)}})^{2\over {4\gamma_0
+1}}\nonumber\\
\nonumber\\
<x>^{1\over 4} = ({2 \eta^{{1+ 2\gamma_0}\over{1+ 4\gamma_0}} \over {h_0 
\lambda (5+ 12 \gamma_0))v_1^{\gamma_0}}}) + ......
\label{v1x1}
\end{eqnarray}
The requirement that $X,Y$ decouple as $m\rightarrow \infty $ amounts to requiring that
$\theta > 10/3 $ and $ \omega > 5$ .
Thus $X$  vanishes as $m\rightarrow \infty$ provided $ -1/4 < \gamma_0 < 1/8$ and $Y$
vanishes if $-1/4 <\gamma_0 <0$  i.e both  decouple only if  $-1/4 <\gamma_0 <0$ .

The values $\gamma_0 =0, -{5\over {12}} $ are seen to be special from eqn.(\ref{v1x1}) and
need separate treatment. In these cases the next to leading order exponents of G are
important :
\begin{equation}
G(v)_{v \to \infty}\longrightarrow g_0 v^{\gamma_0} + g_0' v^{\gamma_0'} + ...
\end{equation}
\vspace{.3 cm}
i) $\gamma_0=0$ , $\gamma_0' > \gamma_0$, by definition, and one finds
\begin{eqnarray}
\delta &=&{{2}\over {1 + 2\gamma_0'}} \Rightarrow \gamma_0' > -{1\over{2}}\nonumber \\
\theta &=&4 \\
\omega &=& 6- {\delta \over 2} \nonumber
\end{eqnarray}

Thus $X,Y$ both decouple as $m\rightarrow \infty$ and the decoupling connection to the pure
$SU(2)$ gauge theory left behind fixes $g_0$ (see the next section).

\vspace{.3 cm}
ii) $v_0=-{5\over {12}}$  Now one finds

\begin{eqnarray}
\delta &=& {{12}\over {1 + 12\gamma_0'}} \Rightarrow \gamma_0' > -{1\over{12}}\nonumber \\
\theta &=& {4 \over {1 + 12 \gamma_0'}} \\
\omega &=& 0
 \nonumber
\end{eqnarray}

Although $X$ decouples for $1/60 > \gamma_0'$,    $Y$ diverges as $m\rightarrow \infty$
so that symmetry is broken completely and strongly .
\vspace{.3 cm}

{\bf{c) $v_0=\infty$}}

\vspace{.3 cm}

i) Here there is again the special case when the leading term in $G(v)$ is a constant:

\begin{equation}
G(v) \longrightarrow  g_{\infty} + g_{\infty}' v^{\gamma_{\infty}'}
\end{equation}

Since now  $\delta < 0$  for consistency  one finds

\begin{eqnarray}
\delta &=& {2\over {1+2 \gamma_{\infty}'}} \Rightarrow \gamma_{\infty}' < -{1\over
2}\nonumber
\\
\theta &=& 4 \\
\omega &=& {{5 +12 \gamma_{\infty}'}\over {1 + 2 \gamma_{\infty}'}}
\nonumber
\end{eqnarray}

Now both the vevs and thus $\Phi$ decouple since $\gamma_{\infty}' < -1/2 $ .
The Konishi anomaly fixes $g_{\infty}$.

\vspace{.3 cm}

ii)  In the generic case when $\gamma_{\infty} \neq 0 $ we get essentially the same equations
as in
{\bf{b)}} above but with $\gamma_0$ replaced by $\gamma_{\infty}$ so that
$\gamma_{\infty} < -1/4$ for consistency.

\begin{eqnarray}
\delta &=& {2\over {4\gamma_{\infty} +1}} \Rightarrow \gamma_{\infty} < -{1\over
4}\nonumber\\
\theta &=& 4 (1-\gamma_{\infty} \delta)\\
\omega &=& {{5 + 12 \gamma_{\infty}}\over {1 + 4 \gamma_{\infty}}} \nonumber
\end{eqnarray}

Since $\gamma_{\infty} <-1/4$ neither $X$ nor $Y$ decouple.

\vspace{.3 cm}
{\bf{II : Perturbative Mode Solutions }}
\vspace{.3 cm}

{\bf{a) $\Phi\neq 0 $ }}
\vspace{.1 cm}
When $\Phi,W_{x,y} $ are non-zero we can multiply eqn(\ref{emphi}) by $\Phi,\Phi^2$ and
take traces to get (using the identity $tr \Phi^4 = (tr\Phi^2)^2/2 $ which follows
using the Jordan canonical form of $\Phi$ )

\begin{eqnarray}
2 x W_x + 3 y W_y =0 \nonumber \\
2 y W_x +{{x^2}\over 2}  x^2 W_y  =0
\end{eqnarray}

so that
\begin{equation}
\langle v\rangle  ={{\langle x \rangle^3} \over {6\langle y \rangle ^2}} =1
\end {equation}

This constraint implies that in the Perturbative mode $\Phi$ must
take a $U(1)$ preserving form which can be taken to be $\Phi =
{\tilde a} Diag (-1,-1,2)$ without loss of generality. Then
$x=6a^2, y=6a^3 = x^{3/2}/{\sqrt 6}$ (where $a={\tilde a} /\Lambda
$) and  the equation for $x$  becomes simply

\begin{equation}
x ( {{\lambda }\over {\sqrt 6}}x^{1/2} +  {{5 G(1)}\over 2}x^{1/4} - \epsilon ) =0
\end{equation}

Since $x=0$ is an artifact of the multiplication by $\Phi$ the solutions are

\begin{equation}
x^{1\over 4} = {{5{\sqrt 3} G(1)}\over {2{\sqrt 2} \lambda}}
\Big [-1 \pm {\sqrt{1 + {{16\eta}\over {25G(1)^2{\sqrt {6}}}}}}\Big ]
\end{equation}

which yields a Small solution (exact if $\lambda =0 $)

\begin{equation}
x_S= ({{2\epsilon}\over {5 G(1)}})^4 + ...
\label{persmall}\end{equation}

and a Large  solution

\begin{equation}
x_L= ({{5{\sqrt 3} G(1)}\over {{\sqrt 2} \lambda}})^4 + ...
\label{perlarge}
\end{equation}

\vspace{.3 cm}
{\bf{b) $\Phi= 0 $ solutions ? }}
\vspace{.3 cm}

In standard Susy GUT symmetry breaking the tree level superpotential allows several
vacuum solutions , among which the trivial or symmetry preserving one is always included.
In presence of a dynamical superpotential however the existence of such a solution becomes
moot. Consider the effect of a typical term $h_{\gamma} v^{\gamma}$ in the series expansion
of $G(v)$ in some region of the v plane. One finds that its contributions to the equations of
motion (\ref{emphi}) are  of form

\begin{equation}
c_1 {{ X^{3\gamma +{1\over 4}} \Phi}\over {Y^{2\gamma}}}
+ c_2 {{ X^{3\gamma +{5\over 4}} \Phi^2}\over {Y^{2\gamma +1}}}
+ c_3 {{ X^{3\gamma +{9\over 4}}}\over {Y^{2\gamma +1}}}  + ....
\end{equation}

\noindent where the $c_i$ are some constants .
Since $\Phi=Diag(a,b,-(a+b) $ ,  $X=2(a^2 + ab + b^2), Y=-3ab(a+b)$, it is clear that
such terms are indeterminate at $a=0=b$ and can be made to take any value depending on how
$a,b$ are sent to zero. Thus it is generically likely that if there is a dynamical
superpotential the $\Phi =0$ solution will no longer exist. We shall see that a similar
conclusion is urged by the Konishi anomaly eqn(\ref{kar}) and the requirement of  correct
decoupling behaviour. In the next section we  discuss the interpretation of these  vevs in
the light of decoupling and the Konishi Anomaly with a view to building DSB models .

\section{Interpretation of Condensates}

Extremization of the superpotential has shown that the dynamically induced superpotential can
cause partial or  complete gauge symmetry breaking  at either a Small scale that
vanishes in the decoupling limit ($m\rightarrow \infty , \Lambda_d, g_d(m_t)$  constant)
or at a Large scale which diverges as  $m,\Lambda \rightarrow \infty  $ with $\Lambda_d$ or
$g_d$ constant. For the Small solutions the entire Higgs multiplet decouples leaving behind an
effective $SU(2)$ N=1 Susy  pure YM theory whose  gaugino condensate is known
to be one of $\langle S\rangle =\pm \Lambda_d^3 $\cite{ads,finpou}. Thus the value of
the $SU(2)$ condensate in the full theory (obtained using the Konishi Anomaly
relation) should match with the pure gauge theory value in the decoupling limit.
This yields constraints on the values  $G(v_0),G'(v_0)$ of the unknown function
$G(v)$ and its derivatives at the decoupling limit values  $v_0$. At finite
values of $m$ the Small solutions could be used
in AS models yielding DSB of gauge symmetries at a {\it{low}} scale e.g for EW breaking
\cite{wip}. However in the case of ASGUT models it is the Large solutions with the
{\it{exponentially}} high scale $\Lambda $ of DSB which are of interest. Note
however that we also obtained solutions where one of the two moduli vanished in
the decoupling limit while the other did not. In the present case the model is
too simple for such behaviour to correspond  to partial symmetry
breaking. However in general when we have many moduli of different types (FM Higgs , AM Higgs
and mixed invariants ) we could have a hybrid Large-Small solution which could be interpreted
as a large breaking of GUT symmetry and a small breaking of the SM symmetry.

For Large solutions, in the complete
symmetry breaking case, the gauge supermultiplet and the entire {\bf{5}}-plet are
supermassive  so that there are no residual degrees of freedom at low energies.
Thus there is nothing to match.
In the  partial symmetry breaking case $SU(2)\rightarrow U(1)$, however,
there is a massless U(1) gauge supermultiplet left over after the decoupling
of the massive charged gauge supermultiplet $W^{\pm}$ with masses
$g\Lambda/\lambda^2 $ and the Higgs residuals
\begin{eqnarray}
\Phi_{\pm 2} &=& {1\over {\sqrt 6}} (\Phi_{11} -\Phi_{22} \mp 2i\Phi_{12} )\nonumber \\
\Phi_0 &=& {1\over{\sqrt 2}} (\Phi_{11} + \Phi_{22})
\end{eqnarray}

\noindent which get masses $\sim \Lambda/\lambda $ (the exact value depends on $G'(1)$).
Since the pure Susy $U(1)$ theory will not condense, matching of the condensates indicates
that the little group ($H$) singlet combination  ($ \langle  \lambda^{\bar a}
\lambda^{\bar a}\rangle $) of the  gaugino
condensate  $\langle \lambda^a \lambda^a\rangle  = \langle  \lambda^{\bar a} \lambda^{\bar
a}\rangle \oplus \langle\lambda^{\hat a} \lambda^{\hat a}\rangle ( {\bar a} \in G/H,
{\hat a} \in H$) should obtain a condensate (if any)  on the scale of the low energy
theory's condensate i.e $0$ for the case of $U(1)$. Since the value of the full
condensate $ \langle  \lambda^{ a} \lambda^{a}\rangle $ is $\sim \Lambda^3$ it seems likely 
that it is the H singlet combinations  $ \langle\lambda^{\hat a} \lambda^{\hat a}\rangle $ (i.e
$ \langle\lambda^+ \lambda^- \rangle$ in the present case)  that will carry the burden of
the large condensate  needed to match the $H$ singlet vevs of the Large Higgs condensates
while the light neutral gauginos  condense only on the scale appropriate to their masses. 
Thus, by analogy,  in the GUT case, it will be  $X$-gaugino bilinears that will
condense to match the Large AM channel  vevs that break the GUT symmetry to $G_{123}$.
Below we discuss the decoupling behaviour of the various Small and Large
solutions obtained by us in detail.

\eject
 
{\bf{Small solutions :}}
\vspace{.2 cm}
{\bf{A) Singlet Mode Solutions :}}
\vspace{.2 cm}

{\bf{ii) $v_0= 0$}}
\vspace{.2 cm}

a) If $\gamma_0\neq 0,-5/12$ we saw that $X,Y$ decouple when $ -1/4 < \gamma_0 <0 $ . However
one finds that

\begin{equation}
\langle S \rangle = {{\Lambda^3}\over {20}} (\epsilon \langle x\rangle -\lambda \langle y
\rangle ) \sim \Lambda_d^3 \eta^{-4\gamma_0\delta}
\end{equation}
Since $\delta, -\gamma_0$ are positive in the decoupling range,  $\langle S\rangle$ will
vanish even though $\Phi$ decouples . Thus this range of exponents is unacceptable
since it does not connect to the known $SU(2)$ gaugino condensate.

b) $\gamma_0=0$
\begin{equation}
\langle S \rangle  =\Lambda_d^3 ({{2^2}\over{5^5g_0^4}} (1 +\quad
O(\eta^{1-{{\delta}\over 2}})))
\label{kar3}\end{equation}

\noindent which is acceptable and fixes $g_0=G(1)$ (the value determined in the perturbative
decoupling solution, see below).

\vspace{.2 cm}
{\bf{iii) $v_0=\infty$}}
\vspace{.2 cm}

The generic solutions with $\gamma_{\infty} <-{1\over 4}$ are non decoupling but for the
special case $\gamma_{\infty}=0$ one obtains decoupling solutions if $\gamma_{\infty}' <
-1/2$ . Evaluating $\langle S \rangle $ one finds

\begin{equation}
\langle S \rangle =\Lambda_d^3 ({{2^2}\over{5^5g_{\infty}^4}}
(1 + O(\eta^{{2\gamma_{\infty}'}\over {1+2\gamma_{\infty}'}})))
\label{kar4}\end{equation}
this decoupling behaviour is acceptable and once again $g_{\infty}=G(1)$.
Note that $\gamma_{\infty}=0$ corresponds to $W\rightarrow x^{5/4}\Lambda^{1/2} +...$ as
$X\rightarrow \infty, Y$ fixed or $Y\rightarrow 0, X$ fixed.

\vspace{.2 cm}

{\bf{B) Perturbative mode solution }}

\vspace{.2 cm}

The vev pattern in this mode, $\Phi \sim Diag(1,1,-2)$, is
preserved by the $T_3$ generator
of $SU(2)$ since $\delta \langle\Phi_{ij}\rangle \sim (\phi_i -\phi_j)\epsilon_{ijk}
\theta_k$ (where $\phi_k , \theta_k , k=1,2,3$ are the diagonal components
of $\langle\Phi\rangle$  and
parameters of $SU(2)$ respectively). In more general situations we expect  that
this type of behaviour will allow for breaking to non-abelian little groups .

The Konishi anomaly relation eqn.(\ref{kar}) gives for the small perturbative solution
eqn(\ref{persmall}) (which decouples leaving $SU(2)$ unbroken as $m\rightarrow
\infty$)  \begin{equation} \langle S \rangle = {\Lambda_d^3\over{20}}
({{ \langle x\rangle}\over \epsilon^4} - {{ \lambda \langle y\rangle}\over \epsilon^5})
=\Lambda_d^3 ({{2^2}\over{5^5G(1)^4}} (1 + O(\eta))
\label{kar2}\end{equation}

\noindent while the value for a pure $SU(2)$ gauge theory is $\pm \Lambda_d^3$ in the DR
scheme \cite{sei1,finpou}. Thus  $G(1)$ is fixed to be

\begin{equation}
 G(1) = ({{\pm 2^2}\over {5^5}})^{1/4}
\label{G1}
\end{equation}

\vspace{.2 cm}
{\bf{b)  Vanishing vev ?  }}
\vspace{.2 cm}

      If the equations of motion were to have the solution $\Phi=0$ we see from
eqn.(\ref{kar}) that even in the presence of a tree superpotential the gaugino condensate
would vanish . Thus the low energy theory which should be pure Susy  SU(2) gauge theory
would not have the condensate it is known to have. Therefore it is natural to expect that
the induced superpotential is such as to destroy the tree level trivial solution. As
discussed above, this is in accord with our expectations from the study of the
equations of motion.

 This behaviour may be compared with a superficially
similar model namely SU(3) Susy YM with one adjoint
chiral superfield . There  all moduli higher than quadratic are  thought to
vanish \cite{rabyank} but   the quadratic modulus (analogous to X) is constrained to
take a value $\sim \Lambda^2$.
However the single adjoint case is rather exceptional in its symmetries.
The $U(1)_0$ R-symmetry under which the  moduli have charge 0 is anomaly free and  the
exponent of $\Lambda$ expected in the superpotential i.e $b_0/(N-T) $  is formally
infinity,  while the charge of $\Lambda^{b_0}$ vanishes. Thus no
dynamical superpotential respecting the non anomalous R-symmetry
can be written in terms of the chiral moduli alone, although the introduction of a Lagrange
multiplier allows one to rewrite the constraint as a equation of motion from a
superpotential (the multiplier is essentially the glueball field $S$).  When further adjoints
are included a superpotential - being permitted
- does arise \cite{rabyank}.  In our model  however the situation is analogous to the
case with several adjoints and one expects the superpotential that can exist to
implement the Konishi anomaly and decoupling.

\vspace{.4 cm}
{\bf{Large Solutions :}}
\vspace{.2 cm}

{\bf{i) $v_0\neq 0,\infty$}}
\vspace{.2 cm}

The symmetry is generically
completely broken although it is possible that
the unknown function $G(v)$ will be such that the singlet mode equations(\ref{singmod})  do
in fact have $v=1$ as a solution( and thus unbroken $U(1)$). In that case one
could match the large solution
of the singlet mode to the large solution in the perturbative mode at $v=1$. The equality
of the values of the leading terms of the perturbative and singlet mode
solutions  at $ v=1$ (see below) is thus  suggestive.

\vspace{.2 cm}
{\bf{ii) $v_0= 0$}}
\vspace{.2 cm}

a) For $\gamma_0 >0 $ the symmetry breaking is Large and complete and there are  no
decoupling constraints .

b) For $\gamma_0=-5/12$ one finds that $Y$ does not decouple since $\omega =0$ so $\langle
S\rangle \sim\Lambda^3 $. Symmetry breaking is again complete although since $X\rightarrow
0$ one can find the explicit form of $\Phi \sim Diag (1, (-1 + {\sqrt 3} i)/2, -(1+{\sqrt
3})/2) $ , which we mention with an eye on the possibilities, by analogy, in deciphering
the GUT case.

\vspace{.2 cm}
{\bf{iii) $v_0=\infty$}}
\vspace{.2 cm}
The generic solutions with $\gamma_{\infty} <-{1\over 4}$ are non decoupling.

{\bf{A) Perturbative Mode Large Solution :}}
\vspace{.2 cm}

In this case since $G(1)$ has been fixed by eqn(\ref{G1}) one finds

\begin{equation}
x_L={{\pm 9} \over {5\lambda^4} } + O(\epsilon) \qquad
 y_L={{x_L^{3\over 2}}\over {\sqrt 6}}
\qquad \langle S\rangle =  -\Lambda^3 {{\lambda x_L^{3\over 2}}\over {20\sqrt 6}}
\end{equation}

Since a U(1) gauge symmetry is unbroken the charged gaugino condensate $\sim \langle
\lambda^+\lambda^- \rangle $ is large  while the light neutral $\lambda^0$ does not condense.

To sum up , the possible types of solutions divide naturally into
Large and Small solutions. The former typically have $\Phi\sim
\Lambda^{\rho}\Lambda_d^{1-\rho} $ while the latter are typically
$\sim (\Lambda_d^3/m)^{\rho'/2}m^{1-\rho'}$ where $\rho, \rho'$
will be determined in practice by the -at present\cite{vafdca}-
unknown function $G(v)$. Irrespective of $G$,
 there is always a $U(1)$ symmetric
solution for which $\rho,\rho'$ are both one. Both kinds of
solutions are constrained to obey decoupling and this has provided
some restrictions on both exponents and amplitudes as well as an
understanding of how the little group and coset sectors of the GUT
scale gaugino condensates should arrange themselves to match the
low scale condensates, if any,  of the effective little group
gauge theory.  These solutions offer an ample range of behaviours
to motivate a study of realistic GUT scenarios of the type
suggested in \cite{trmin} to determine the dynamical symmetry
breaking possibilities. Indeed, Susy AS YM theories can even be
considered in other important contexts such as novel types of DSB
models for EW symmetry breaking\cite{wip} by employing the Small
solutions . In the next section we discuss the gross features of
prospective ASGUTs in the light of what we have learnt from our
toy model.

\section{Realistic AS GUTs}

As discussed in detail in \cite{trmin}, our primary motivation for
studying ASGUTs - apart from their theoretical fascination as the
`dark side'  of YM theories - is that {\it{typical $SU(5)$ and
$SO(10)$ Susy GUTs that successfully account for realistic charged
fermion and neutrino masses
 at the tree level (i.e without invoking Planck scale effects and non
renormalizable operators) are Asymptotically Strong }}.   A realistic
 $SU(5)$ model should consist (at least ) of
three families of matter in ${\bf{\bar 5}}(F)\oplus {\bf{10}}(T)$
, FM Higgs  ${\bf{5}}(H)\oplus {\bf{{\bar 5}}}({\bar H})$  and $
{\bf{45}}(\Sigma)\oplus {\bf{\overline{45}}}({\bar\Sigma}) $
(which drive the gauge coupling strong above the   thresholds
($\sim M_U$) of the massive residual FM Higgs sub-multiplets and allow one to fit fermion
masses \cite{realmf1,realmf2}) and an AM Higgs
${\bf{24}}(\Phi_i^j$) to enable us
to describe the GUT symmetry breaking as a classical condensate.

The renormalizable tree superpotential in this  model is quite
restricted. It further simplifies once one imposes  R-parity.
R-parity is central to our approach to Susy
unification \cite{abs,ams,amrs,rpuni} since it defines and
maintains the distinction between matter fields and vacuum
defining order parameter fields , which one  violates only at the
risk of fancifulness (as opposed to minimality )  in view of the
stringent constraints available and the lack of  direct evidence
for supersymmetry. Then only operators with an even number of
matter superfields are allowed in the tree superpotential. The
possible gauge invariants (up to cubic) are (flavour indices are
suppressed and the masses /couplings
  included in the definitions for
convenience) :

\begin{eqnarray}
Z_1 &=& m_F \Sigma {\bar{\Sigma}} \qquad  Z_2 = {{m_A}\over 2} \Phi^2  
\qquad Z_3= {\lambda \over 3}
\Phi^3
\nonumber \\
Z_4 &=& w_1  \Sigma {\bar{\Sigma}}\Phi   \qquad Z_5 =  y_1 T{\bar F} {\bar\Sigma } \qquad
Z_6 =   {{y_2}\over 2} TT\Sigma   \\
Z_7 &=&  m_H H {\bar{H}}  \qquad Z_8 = {{y_3}\over 2} TTH  
\qquad Z_9= w_2 {\bar{H}}\Phi H  \nonumber\\
Z_{10} &=& y_4 T{\bar F}{\bar H}  \qquad Z_{11} = \theta_1 {\bar\Sigma}H \Phi
\qquad  Z_{12} = \theta_2 {\Sigma}{\bar H} \Phi
\end{eqnarray}
 with obvious contractions.
Then the renormalizable R-Parity preserving superpotential is simply $W_t=\sum_i Z_i $.

The construction of the possible form of the dynamical superpotential
is now a very complex project
since the number of chiral invariants even in the $\Phi-\Sigma-{\bar \Sigma}$ sector is very
large. We therefore restrict ourselves here to considering how the symmetry breaking
in a successful model  might proceed and the restrictions on the pattern of
condensates discernable on the basis of the  Konishi anomaly.

Let us consider the problem in stages . We first retain only the
$\Phi-\Sigma $ sector i.e invariants $Z_1 ...Z_4$.
The Konishi anomaly gives (we drop $\langle \rangle $  for brevity)

\begin{eqnarray}
-24 S &=& Z_1 + Z_4 \nonumber\\
-10 S &=& 2  Z_2 + 3 Z_3 + Z_4
\label{phisig}
\end{eqnarray}

With 4 invariants and the gaugino condensate we see that the
 two equations leave 3 invariants (say $Z_1,Z_2,Z_3$)
free while

\begin{eqnarray}
Z_4 &=& w_1\Phi \Sigma {\bar\Sigma} = {1\over 7}(5 Z_1 -24 Z_2 -36 Z_3) =
{5\over 7} m_F \Sigma {\bar{\Sigma}} -{{12}\over 7} (m_A \Phi^2 + \lambda \Phi^3)
\nonumber \\
S  &=& {1\over {14}} (-Z_1 + 2 Z_2 + 3 Z_3) =
{1\over {14}} (- m_F \Sigma {\bar{\Sigma}}  + m_A \Phi^2 + \lambda \Phi^3)
\label{solphisig}
\end{eqnarray}
\noindent In our scenario the vevs have magnitudes  $\Phi \sim \Lambda ;
 \Sigma, {\bar \Sigma} \sim M_W $ while $ m_F \sim M_U$ , thus it is clear that this
constraint can only be satisfied either if the combination
$(m_A \Phi^2 + \lambda \Phi^3)$ is vanishing to $O(M_W^2)$ 
(either from the solution itself or by fine
tuning) or else the fields are living in a Quantum vacuum where  $\Sigma {\bar{\Sigma}}$
attains a large vev but  the fields $\Sigma, {\bar{\Sigma}}$ have vevs $\sim M_W $.
At least with present calculational techniques,
the latter alternative represents a regression to the pre-calculable or ad-hoc stage
as far as the effective theory of interest is concerned. The price of the first
illustrates the adage concerning  free lunches !

To proceed further let us now add a single family of matter fields
i.e include the two additional invariants $Z_{5,6}$.
The KA equations are now
\begin{eqnarray}
-S &=& Z_5\nonumber\\
-3S &=& Z_5 + 2 Z_6 \nonumber\\
-24 S &=& Z_1 + Z_4 +  Z_5\nonumber\\
-24 S &=& Z_1 + Z_4 +  Z_6\nonumber \\
-10 S &=& 2  Z_2 + 3 Z_3 + Z_4
\label{phisigFT}
\end{eqnarray}
whose solution is :
\begin{eqnarray}
Z_4 &=& w_1\Phi \Sigma {\bar\Sigma} = {1\over {13}}(10 Z_1 -23(2 Z_2 +3 Z_3)) =
{1\over {13}} \big(10 m_F \Sigma {\bar{\Sigma}} - 23 (m_A \Phi^2 + \lambda \Phi^3) \big )
\nonumber \\
Z_5 &=& Z_6= -\langle S \rangle  = {1\over {13}} (Z_1 -( 2 Z_2 + 3 Z_3)) =
{1\over {13}} ( m_F \Sigma {\bar{\Sigma}}  -( m_A \Phi^2 + \lambda \Phi^3))
\label{solphisigFT}
\end{eqnarray}

\noindent While the argument concerning fine tuning still holds for the
 $\Phi,\Sigma,{\bar \Sigma} $ condensate relations
we see that some new feature must intervene to avoid invoking `` quantum vacua''
if we are to avoid catastrophic violation of Baryon and Lepton
 number through vevs of the sfermions in the matter
{\bf{5,10}} ! Although we can invoke the violation of supersymmetry and thus of the
Konishi Anomaly relations (\ref{kar}) by terms $\sim M_S^3 $ a
further fine tuning to
ensure that $S\sim M_S^3$ seems required. An even more rigorous
fine tuning is required if one insists that each
equation be modified only by terms $\sim {M_S^{i}}^2 \langle \phi_i\rangle $
 since  the $F,T$ vevs must
vanish very strictly. A decision on the consistency of the former
choice requires a review of the derivation of the
Konishi anomaly in the presence of soft supersymmetry breaking.
The possibility that the vevs of the composite
invariants involving matter superfields are non zero while the vevs
of the fields themselves
are not also deserves further consideration , particularly since it
would contribute novel features to the effective
theory (analogous to instanton-chiral fermion vertices but now with chiral scalars 
disappearing into the vacuum in gauge and B-L invariant combinations).

One can continue in this vein, adding the invariants $Z_7...Z_{12}$ containing $H, {\bar H}
$, but we omit the details as the point regarding the necessity of fine tuning have already
been made clear.

Although many details need to be worked out we see that so far the Truly Minimal(TM)
ASGUT program has not run into 
manifestly insuperable obstacles of interpretation or consistency and
 can even furnish a low energy
effective theory largely  identical to that of standard Susy GUTs with ,
 however, a radically modified picture of the root dynamics that engendered it and a
way of generating the exponential gauge hierarchy of GUT
theories that is intrinsic to the RG running that suggested it's existence in the first place 
\cite{gqw}.
 Since we are very far from verifying any feature of the
standard Susy GUT dynamics at the unification scale it would be premature to
foreclose the option of Asymptotic strength.
In the next section we shall see that ASGUTs offer further novel
insights into elementarity and a qualitatively new cosmogony.

\section{ASGUTs, Dual Unification and Concluding Remarks}
\tighten
Our new picture (Fig.1) of an elementary particle bears an intriguing duality to the
picture of standard model particles as monopoles formed in the breaking of a dual
unification group down to a dual standard model \cite{vachas}. In both pictures
elementary particles have a core of very small size :
$l_m \sim {\widetilde M}_U^{-1} $ in the dual unification
picture and $ l_e \sim \Lambda_U<<< M_{W}^{-1}$ in the
ASGUT picture . The monopole core of Higgs/gauge magnetic
energy is naturally dual to the tangle of electric flux that forms at the core of the
 elementary particle in the AS  theory  with both representing concentrations
of mass-energy of utterly tiny size. Thus it is very natural to identify
$\l_m\sim l_e$ i.e $  {\widetilde M}_U \sim \Lambda_U $ .
We argue below that  ASGUTs present ready answers to
some of the difficulties faced by  dual unification models
when we make this identification of the
dual unification scale ${\widetilde M}_U$   not with the perturbative
electric unification scale $M_U$ but with the UV condensation scale $\Lambda_U$.

Monopole solutions of the classical equations of motion   make sense only in the regimes
where the   gauge coupling  $g_m(l_m)$
is small.  Therefore  the  electric  dual of the magnetic coupling
 $g_e= 2\pi/g_m$ will  be  large at the same energy scale i.e {\it{the electric theory should
be AS}}.    Furthermore the discrepancy
between the large mass one would expect for a weakly coupled  monopole
  with a core of size $\sim  {\widetilde M}_{U}^{-1}$  and the GeV scale masses
 of  SM particles (at the EW scale, or even -after renormalization -
 at the perturbative unification scale $M_U$) receives an elegant explanation
 in terms of the rapid growth of the effective mass of SM non singlet
 masses with the growth of the gauge coupling and the vanishing of
their wave function renormalization constant as the confinement
 scale $\Lambda_{U}$  lying above $M_{U}$  is approached : the
huge dual model soliton mass at ${\widetilde M}_U$ can thus be
identified with the large mass of the gauge variant
particles as they approach confinement.
Correspondingly the dual model's  monopole mass $\sim {\tilde v}/g_m$
will  fall as its coupling grows strongly below $\Lambda_{U}$
to match the rapidly falling  electric coupling . In other words the
dual identification ${\tilde \Lambda_U} \sim M_U $ is also naturally implied.
   This picture is also in accord with the large values of the magnetic
coupling that may be expected on the basis of the known values of the
 couplings of the SM and the duality relation $4 \alpha_m \alpha_e =1 $. These
show that the dual gauge group may be expected to be confining at
all energies below  ${\tilde \Lambda}\sim M_U $ corresponding to
confinement of its particles and of monopoles of standard GUTs and
providing a clear reason for their non observation. {\it{ This constitutes a novel and robust
solution of the monopole problem\cite{preskill} of GUTs.}}

Finally, in the context of a Supersymmetric AS GUT  one enjoys the additional
bonus of complete supermultiplets of dual monopoles
dual to the chiral supermultiplets containing the fermions and sfermions of the  MSSM.
This avoids the necessity of conferring a half integer spin on the bosonic monopoles by
delicate additional  mechanisms \cite{lykvach} : both quarks and squarks arise
together. Since the dual model is AF, these arguments motivate a supersymmetric
dual unification scenario \cite{wip} using e.g an AF $\widetilde{SU(5)}$ gauge 
theory with adjoint Higgs broken to ${\tilde G}_{123}$ at a scale 
${\widetilde{M_U}}\sim \Lambda_U$.

The high temperature behaviour of AS Theories and in particular ASGUTs is
also remarkable. One expects that the growing gauge coupling at high temperature will
cause all the  gauge variant  fields to condense and their fluctuations
will acquire infinite effective masses leaving behind only a few gauge singlet degrees of freedom
associated with the D-flat chiral moduli, which typically have masses in the region of
$\Lambda$, to carry the entire energy content of the (cosmological) system. Conversely
{\it{cooling }} below $\Lambda_U$ liberates the gauge degrees of freedom and the gauge
variant fermions and scalars ! It is this pleasing cosmogony in which the ``featureless''
(i.e SM singlet) but ultra-energetic initial plasma freezes out 
into a manifold plurality that inspired the name 
{\it{pleromal \footnotemark}} \cite{trmin}. These considerations
motivate a detailed study of the pleromal gauge singlet dynamics at high temperatures.
This may be undertaken using standard Susy sigma model techniques.

\footnotetext{1) Pleromal [{\it{a. Gr.}} $\pi\lambda\eta\rho\mu\alpha$  that which
fills],Fullness, plenitude; In Gnostic theology, the spiritual universe as the abode 
of God and the totality of the Divine powers and emanations 2) Plerome: {\it{Bot.}} 
The innermost layer of the primary tissue or meristem at a growing point.
(Shorter Oxford Dictionary)} 
\tighten

The relation of ASGUTs to gravitation is clearly of possible fundamental importance. We
restrict ourselves to remarking here that the common feature of Asymptotic Strength of
Gravity and ASGUTs, together with Susy \cite{david} and the closeness
$\Lambda_U\sim M_{Planck}$ is a plausible motivation for re-evaluating the program of
induced\cite{sakharov,adler} (super)\cite{david}-gravity  in the context of
a Susy ASGUT with background metric and gravitino fields
 in the fundamental lagrangian introduced to
permit the implementation of (super) general
coordinate invariance. As is well known \cite{wewi} this scenario
by passes the no-go theorems limiting the generation of massless
particles with spin greater than  $1/2$ \cite{casgas,wewi}.

A large number of complex dynamical issues must be resolved before the viability of realistic
ASGUT or AS-Technicolour scenarios can be established . Nevertheless our hope
is to have contributed a stimulus to the taking up of this challenge by
breaking the taboo against AS theories by sketching `pretty pictures'  in
which they belie their  reputation of fearsome and untameable
strength.

{\bf{Acknowledgements :}}

It is a pleasure to acknowledge the warm hospitality of the High Energy Group
of the International  Centre for Theoretical Physics, Trieste where this work
was conceived and executed and  Goran Senjanovic in particular for 
hospitality, friendship and useful advice. I am grateful to Gia Dvali,
Hossein Sarmadi, Asoke Sen and Tanmoy Vachaspati for very useful 
discussions and T.J. Hollowood, K. Konishi  and M.Shifman for helpful
correspondance. I thank George Thompson for help with TeX and figures.

\end{document}